\documentclass[twoside,12pt]{article}  
\usepackage{indentfirst}  
\usepackage{bm}    
\usepackage{graphicx}  
\usepackage{graphicx}
\usepackage{dcolumn}
\usepackage{bm}
\usepackage[caption=false]{subfig}
\usepackage{wasysym}
\usepackage{url}

\begin{document}    

\thispagestyle{empty} \vspace*{0.8cm}\hbox
to\textwidth{\vbox{\hfill\huge\sf Chinese Physics B\hfill}}
\par\noindent\rule[3mm]{\textwidth}{0.2pt}\hspace*{-\textwidth}\noindent
\rule[2.5mm]{\textwidth}{0.2pt}


\begin{center}
\LARGE\bf The electronic structure and elastic property of  monolayer and bilayer transition metal dichalcogenides MX$_2$ (M=Mo,W;X=O,S,Se,Te): A comparative first-principles study$^{*}$   
\end{center}

\footnotetext{\hspace*{-.45cm}\footnotesize $^*$This work was  supported by the construct program of the key discipline in hunan province and aid program for Science and Technology Innovative Research Team in Higher Educational Institutions of Hunan Province.}
\footnotetext{\hspace*{-.45cm}\footnotesize $^\dag$Corresponding author. E-mail: zhangwb@csust.edu.cn (Wei-Bing Zhang); tangbiyu@gxu.edu.cn(Bi-Yu Tang) }

\begin{center}
\rm Fan Zeng$^{\rm a,b,c}$, \ \ Wei-Bing Zhang$^{\rm b}\dagger$, \ and  \ Bi-Yu Tang$^{\rm a,c}\dagger$
\end{center}

\begin{center}
\begin{footnotesize} \sl
${}^{\rm a)}$ Department of Physics, Xiangtan University, Hunan Province 411105, People's Republic of China \\   
${}^{\rm b)}$ School of Physics and Electronic Sciences, Changsha University of Science and Technology, Changsha 410114, People's Republic of China \\   
${}^{\rm c)}$ School of Chemistry and Chemical Engineering, Guangxi University, Nanning 530004, People's Republic of China\\   
\end{footnotesize}
\end{center}

\begin{center}
\footnotesize (Received X XX XXXX; revised manuscript received X XX XXXX)
\end{center}

\vspace*{2mm}

\begin{center}
\begin{minipage}{15.5cm}
\parindent 20pt\footnotesize
First-principle calculations with different exchange-correlation functionals, including LDA, PBE and vdW-DF functional in form of optB88-vdW, have been performed to investigate the electronic and elastic properties of two dimensional transition metal dichalcogenides(TMDCs) with the formula of MX$_2$(M=Mo,W; X=O,S,Se,Te) in both monolayer and bilayer structures.  The calculated band structures show a direct band gap for monolayer TMDCs at the K point except for MoO$_2$ and WO$_2$. When the monolayers are stacked into bilayer, the reduced indirect band gaps are found except for bilayer WTe$_2$, in which direct gap is still present at the K point. The calculated in-plane Young moduli are comparable to graphene, which promises the possible application of TMDCs in future flexible and stretchable electronic devices. We also evaluated the performance of different functionals including  LDA, PBE, and optB88-vdW in describing elastic moduli  of TMDCs and found that  LDA seems to be the most qualified method. Moreover, our calculations suggest that the Young moduli for bilayers are  insensitive to stacking orders and the mechanical coupling between monolayers seems to be negligible.
\end{minipage}
\end{center}

\begin{center}
\begin{minipage}{15.5cm}
\begin{minipage}[t]{2.3cm}{\bf Keywords:}\end{minipage}
\begin{minipage}[t]{13.1cm}
 Transition metal dichalcogenides; Bilayer structures;  Elastic properties; Electronic structure; First-principles calculation
\end{minipage}\par\vglue8pt
{\bf PACS: }
71.15.Nc; 62.20.de; 62.20.dj; 73.22.-f
\end{minipage}
\end{center}

\section{\label{sec:Intro}Introduction}

 The successful isolation of graphene has brought in great revolutions for modern materials science $^{[1]}$. The fascinating properties, such as the exceptionally high electron mobility ( $\sim$ 10$^5 cm^2V^{-1}s^{-1}$)$^{[2]}$ and unique combination of high modulus ($\sim $ 1000 GPa) and tensile strength ($\sim$ 100 GPa)$^{[3]}$, have made graphene to be one of the most promising candidates for future flexible and stretchable electronics. However, the lack of band gap in pristine graphene limits its reality application. Due to the presence of a sizeable direct band gap in the visible frequency range $^{[4,5]}$, two-dimensional transition metal dichalcogenides (TMDCs) monolayers have recently suggested as an important candidate for electronic and optoelectronic devices $^{[4,5,6]}$. Meanwhile, TMDCs also exhibit a number of intriguing optical phenomena such as valley-selective circular dichroism and the rich physics associated with the valley degree of freedom $^{[7]}$. Moreover, their band gaps are tunable with thickness and different monolayer TMDCs can also be reassembled into designer van der Waals heterostructures $^{[8]}$, which may lead to even richer physics and device application.

Mechanical property of materials such as elastic modulus is of vital importance in any practical device applications. It is thus a necessary prerequisite for the integration of TMDCs in various devices to obtain a detailed knowledge of mechanical properties of monolayer and 2D heterostructures. Using nanoindentation in an atomic force microscope, Bertolazzi \emph{et al.} $^{[9]}$ have investigated in-plane stiffness and breaking strength of suspended monolayer and bilayer MoS$_2$, which indicates that monolayer MoS$_2$ exhibits exceptional mechanical properties comparable to stainless steel. In-plane stiffness of monolayer and bilayer MoS$_2$ are found to be 180$\pm$60 Nm$^{-1}$ (corresponding to an effective Young's modulus of 270$\pm$100 GPa) and 260 $\pm$70 Nm$^{-1}$(200$\pm$ 60 GPa), respectively. Recently, the elastic modulus of CVD grown monolayer MoS$_2$ and WS$_2$ and their Bilayer Heterostructures are also investigated $^{[10]}$. A high 2D elastic moduli of CVD monolayer MoS$_2$ and WS$_2$ ($\sim$ 170 Nm$^{-1}$) are also found. The 2D moduli of their bilayer heterostructures are found to be lower than the sum of 2D modulus for each layer, which shows a strong mechanical interlayer coupling between the layers $^{[10]}$.

The importance of mechanical properties in future devices application also intrigues theoretical interest. The in-plane stiffness of monolayer MoS$_2$  have been investigated extensively $^{[11,12,13,14,15]}$. However, the scatted theoretical results from  124 to 146 Nm$^{-1}$ seems to be lower than the experimental value of 170 Nm$^{-1}$. Recently, the in-plane stiffness of a series of monolayer MX$_2$ (M=Mo, W; X=S, Se, Te,O) has also been investigated by Kang \emph{et al.} $^{[13]}$ and \c{C}akr \emph{et al.} $^{[14]}$. Their results suggested that TMDCs with W (O) atom are found to be much stiffer in each chalcogenide (metal) group. It should be noticed that the Young modulus of TMDCs obtained using PBE functional seems to be lower than experimental measure as shown above, it is thus interesting to evaluate the performance of commonly-used functionals to describe the mechanical properties of TMDCs. On the other hand, in contrast to the extensively studies for monolayers, elastic and mechanical properties of bilayer TMDCs still lack completely. Previous study has suggested a transition from direct gap in monolayer to indirect gap in bilayer for TMDCs $^{[4, 16, 17, 18]}$, which implies a strong electronic coupling between different monolayers in bilayer TMDCs.  Meanwhile, stacking order is also found to affect the electronic band structure and absorption spectra $^{[19]}$. It is thus quite interesting to investigate mechanical coupling between monolayers and to reveal the role of stacking order on  elastic property of bilayer TMDs theoretically .

To understand electronic and  mechanical  coupling of monolayer in bilayer TMDCs, we have investigated electronic and elastic properties of eight kinds of TMDCs in both monolayer and bilayer structures , in which different functionals are employed for comparison. The detailed structure, electronic properties, elastic constants and Young modulus of  TMDCs both in monolayer and bilayer obtained using LDA ,PBE and optB88-vdW functional are given and compared with available experiment and theory. Moreover, the performance of different functionals in describing the elastic properties of TMDCs is also evaluated.

\section{\label{sec:method}Computational Details}

 The present calculations were performed with the Vienna Ab-initio Simulation Package (VASP) code $^{[20, 21]}$ within the projector augmented wave (PAW) method. The local density approximation (LDA) and Perdew-Burke-Ernzerhof(PBE) are used in our calculation for comparison. To take into account van der Waals forces, which are expected to play a crucial role in bilayer, the van der Waals density functional in form of optB88-vdW $^{[22, 23]}$ was used. Recent study has been evidenced that the optB88-vdW can give much improved results for graphene/metal surface $^{[24]}$. Cut off energy and Brillouin zone (BZ) sampling are determined after extensive convergence test. The Brillouin zone is simpled by a (18$\times$ 18 $\times$ 1) Monkhorst-Pack grid and a plane-wave basis set with kinetic cutoff energy of 550 eV is applied. The atomic positions of those 2D TMDCs have been optimized by using the conjugate-gradient(CG) algorithm. A vaccum  of 20 \AA~ was set to prevent interactions between periodic images.The elastic constants are obtained by strain-energy method. By applying a set of suitable deformations to the single unit cell of the 2D TMDCs, we can get the energy-strain curves, and the corresponding elastic constant can be obtained. The detailed relations between applied strain and elastic constant are also shown in Ref $^{[25]}$. The strains ranged from -0.05 to 0.05 in steps of 0.01 are  used in the present calculation.

\section{\label{sec:results}Results and Discussion}
\subsection{\label{sec:ML} Monolayer TMDCs}
 \subsubsection{\label{sec:ele}Structure properties}

 \begin{figure}
\centering
\includegraphics[width=0.6\textwidth]{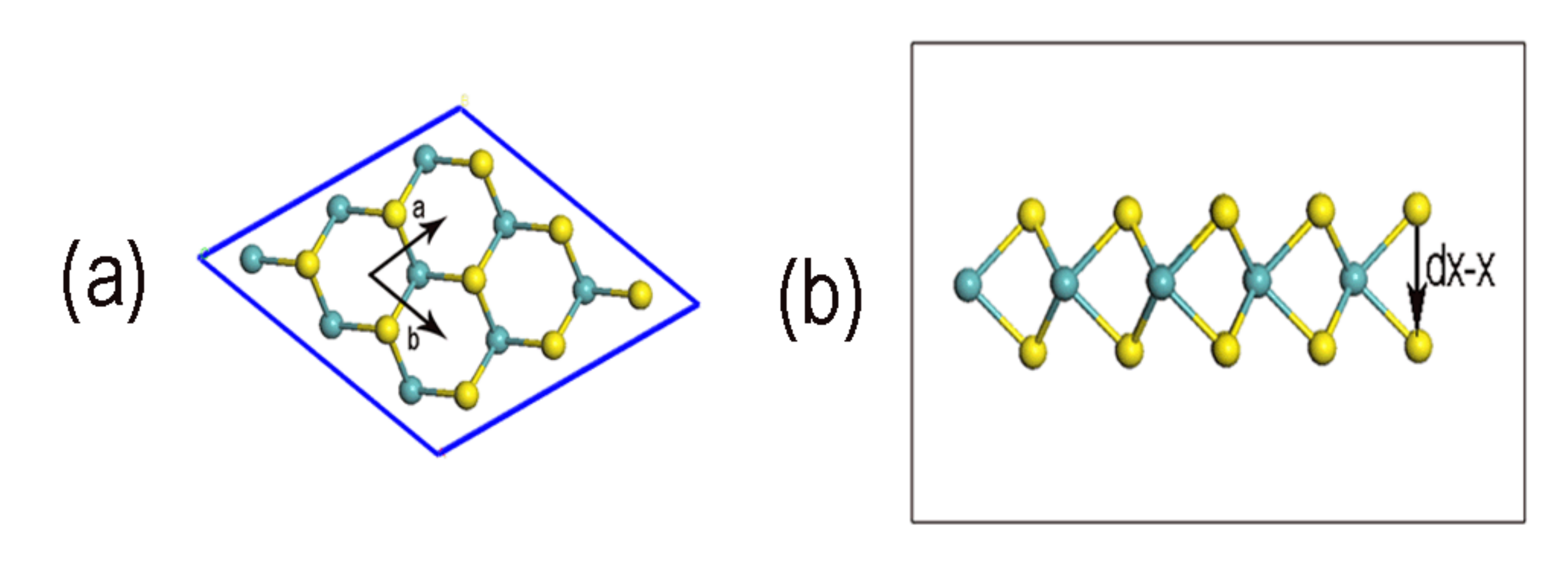}
\includegraphics[width=0.6\textwidth]{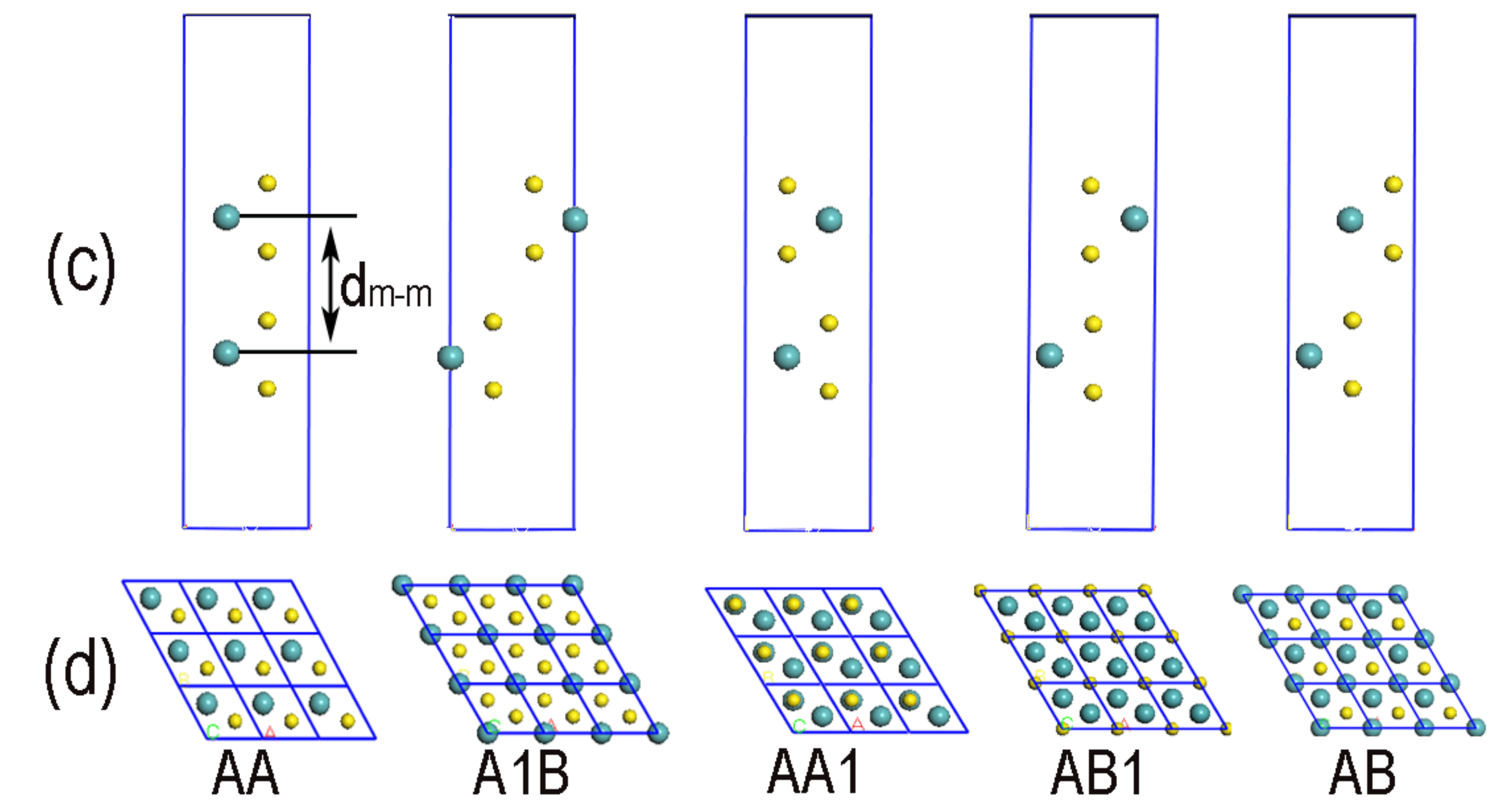}
\caption{\label{fig:geom}(Color online) Top view (a) and side view (b) of the monolayer TMDCs. (c) and(d) are stacking structure for bilayer. Yellow balls and green balls stand for metal and chalcogen atoms respectively. }
\end{figure}

 The structure of monolayer TMDCs  is shown in Fig.~\ref{fig:geom}-(a), in which the M and X atoms occupy the hexagonal honeycomb site alternately. Due to the chemical ratio of M:X=1:2, the M sublattice layer is sandwiched between two nearby X sublattice layers and forms an X-M-X covalently bonded hexagonal quasi-2D lattice as shown in Fig.~\ref{fig:geom}-(b), rather than the planar construction of the graphene.

 \begin{table}
\centering
\caption{\label{tab:ML_geom}The optimized structural parameters of monolayer TMDCs using LDA,PBE and optB88-vdW method.  a and d$_{x-x}$ stand for lattice constant and bond length of two X atoms, respectively. The PBE results from Ref.$^{[13]}$ and Ref.$^{[14]}$  are also listed for comparison. }
\small
\begin{tabular}{cccccccccccccc}
\hline
\hline
	&		&	&a(\AA)		&		&&&	d$_{x-x}$(\AA)	&		\\
	&	LDA	 &PBE	&	optB88	& Ref.$^{[14]}$ &Ref.$^{[13]}$ &	LDA	&	PBE	&	optB88	\\
\hline
MoO$_2$	&	2.79	&	2.83	&	2.83&2.83&-	&	2.44	&	2.47	&	2.47	\\
MoS$_2$	&	3.12	&	3.18	&	3.19&	3.18&3.18&	3.11	&	3.13	&	3.14	\\
MoSe$_2$	&	3.24	&	3.32	&	3.32&3.32&3.32&	3.31	&	3.33	&	3.34	\\
MoTe$_2$	&	3.46	&	3.55	&	3.56	&3.54	&3.55&	3.59	&	3.61	&	3.62	\\
WO$_2$	&	2.79	&	2.85	&	2.83	&	2.83&-&2.44	&	2.48	&	2.48	\\
WS$_2$	&	3.12	&	3.18	&	3.19	&	3.18	&3.18&3.12	&	3.14	&	3.15	\\
WSe$_2$	&	3.24	&	3.32	&	3.32	&3.32	&	3.32&3.33	&	3.35	&	3.36	\\
WTe$_2$	&	3.47	&	3.55	&	3.56	&3.55	&	3.55&3.6	&	3.63	&	3.63	\\
\hline
\hline
\end{tabular}
\end{table}
 The optimized structural parameters of  TMDCs with LDA, PBE and optB88-vdW methods are listed in Table~\ref{tab:ML_geom}, two previous PBE calculations $^{[13, 14]}$ are listed for comparison.  We can find that the previous calculations are almost identical with our PBE results, which suggests that the present results are reliable.  It can be clearly seen that the structure seems to be independent to metal element but increase monotonously with atomic number of chalcogen.  In addition, it can be also found that LDA calculation gives a smaller value than PBE, while the results of PBE and optB88-vdW are almost the same, which are also more close to bulk crystals. This is in agreement with well accepted knowledge that LDA overestimates the binding of materials.    The similar result of PBE and optB88-vdW indicates van der Waals functional can provide a balanced description for systems dominated by not only vdW force but also covalence bonding, and  van der Waals interaction has few effects on such a monolayer TMDCs.

  \subsubsection{\label{sec:ele}Electronic properties}

   \begin{figure}
\centering
\includegraphics[width=0.8\textwidth]{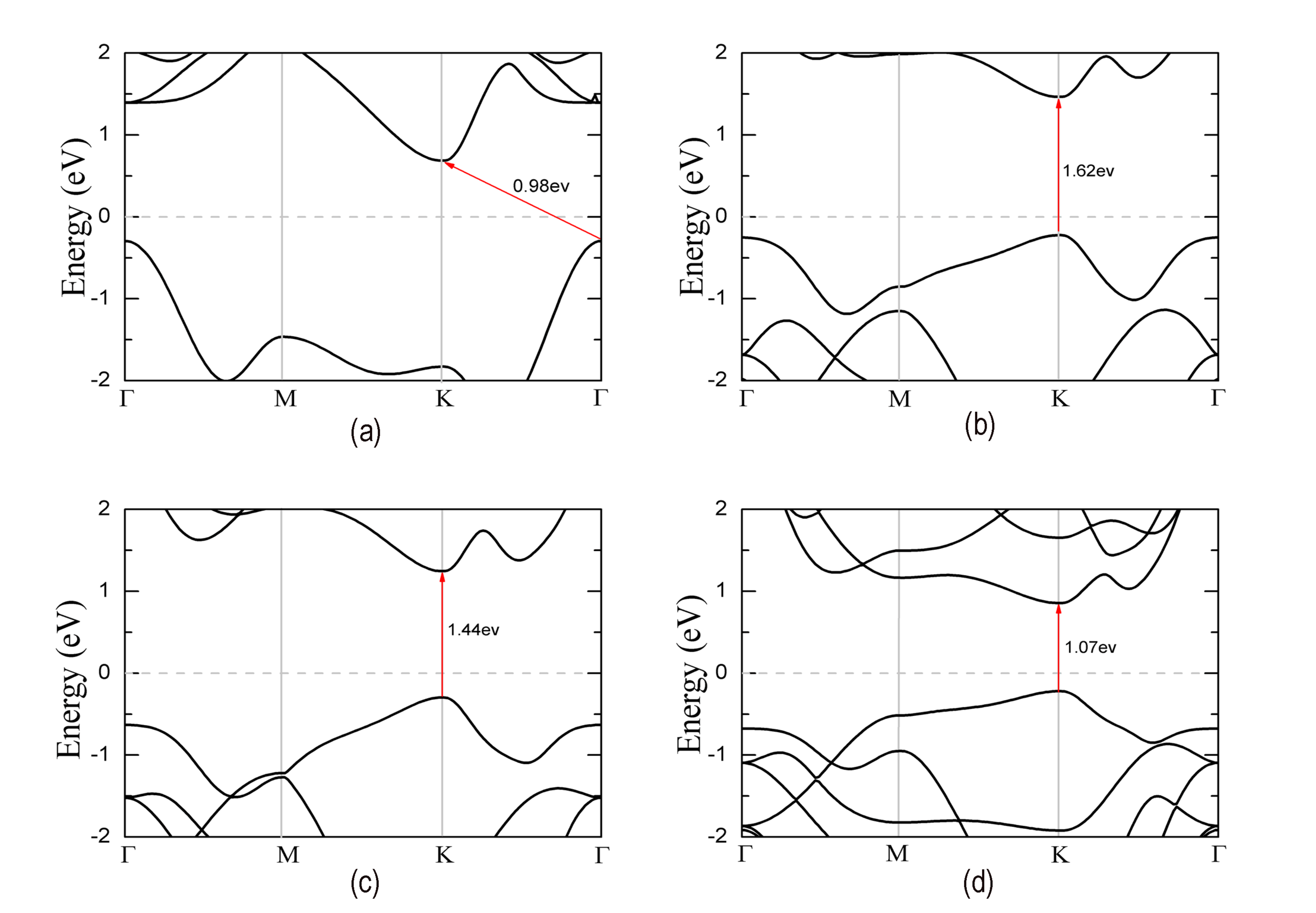}
\includegraphics[width=0.8\textwidth]{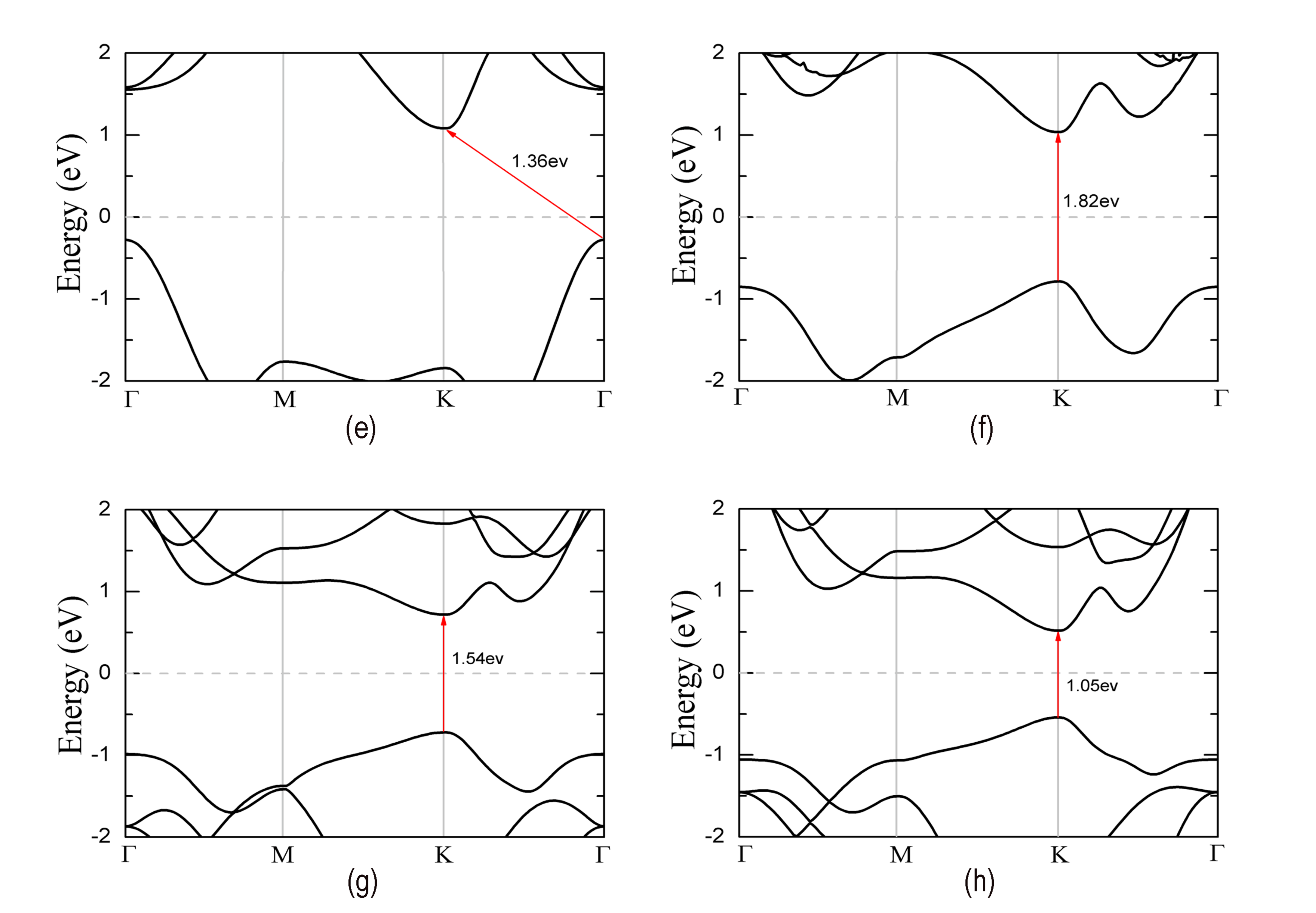}
\caption{\label{fig:ML_band}(Color online) The band structures for monolayer TMDCs calculated using PBE. (a)-(h) stand for MoO$_2$,MoS$_2$,MoSe$_2$,MoTe$_2$,WO$_2$,WS$_2$,WSe$_2$,WTe$_2$ respectively. The Fermi energy is marked with light gray dash lines. The band gaps are shown with red arrows. }
\end{figure}

  Now, let's focus on band structures of monolayer TMDCs. Since the (semi)-local functionals are known to give similar band structure for materials, we only give the energy band of monolayer TMDCs calculated using PBE functional. As shown in Fig.~\ref{fig:ML_band}, all those eight kinds of single layer TMDCs are semiconductors with a distinct band gap, which is quite different from the zero-band-gap graphene. For example, the monolayer MoS$_2$ has a direct band gap at the K point with a band gap value of 1.62 eV, which is close to the experimental study for atomically thin MoS$_2$ layers $^{[4, 5]}$. The direct band gap at the K point can be also found in the case of monolayer MoSe$_2$ and MoTe$_2$, and the corresponding band gap is 1.44 eV and 1.07 eV, respectively. Howerver, for the case of MoO$_2$, the band gap is indirect with a value of 0.98 eV. For monolayer MX$_2$ (where M stand for Mo and W) with direct energy gap, the band gaps decrease when the chalcogenides element becomes heavier \emph{i.e.} the energy gap order is disulfide$>$diselenide $>$ditelluride. To analysis the underlying mechanism, we have performed Bader charge analysis, which is know to be able to get the charge transfer between different atomic species . Taking MoX$_2$ as an example, we can find the total charge of Mo atom in monolayer MoX2 is 4.29, 4.52, and 5.06 e, which suggests that the charge transfer from metal to chalcogen atom is 1.71, 1.48, 0.94 e. This indicates that the reactivity and binding decrease from S to Te, which can also be supported by the fact that the lattice constants decrease with the row number of X in Mo dichalcogenides. Clearly, the strong reactivity of chalcogen atom will enhance the hybridization, which will lead to larger band gap of the compounds. In addition, the band gap for MoX$_2$  (X=Se, S) seems to be smaller than WX$_2$,  and MoTe$_2$ and WTe$_2$ have quite close gap. However, For monolayer oxide with indirect-band-gap, the band gap of WO$_2$ is much larger than that of MoO$_2$. The calculated band gaps agree well with other studies $^{[13, 27]}$.

  \subsubsection{\label{sec:ele}Elastic properties}
  Elastic property is another important physics quantity focused on the present work. For 2D TMDCs, it is possible to measure the elastic constants of these materials from nanoindentation experiments using an atomic force microscope $^{[28]}$. However, since the experimental measurement suffers from some uncertainness such as defect, First-Principle calculation can offer an important supplement to experiment.

  The elastic constants for monolayer TMDCs calculated with PBE method are reported in Table ~\ref{tab:ML_elastic}. Since the monolayers of 2H-MX$_2$ belong to isotropic structures, the linear elastic constant C$_{22}$ is equal to C$_{11}$. Linear elastic constants of MoX$_2$ and WX$_2$ decrease monotonically with the increase of chalcogen's atomic number, which means the decrease of the Young modulus. And for the same chalcogenide atoms, the WX$_2$ compounds have larger elastic constant than MoX$_2$ compounds. The Poisson ratio which is used to describe the lateral deformation is calculated by $\nu=C_{12}/C_{11}$ . It is remarkable that the Poisson ratios of those 2D TMDCs are larger than graphene except for WTe$_2$, which means much better lateral contraction property than graphene. The calculated Poisson ratio for MoS$_2$  is also close to other studies $^{[12, 14]}$.

For further discussion of elastic property of those monolayer TMDCs, we also evaluate the Young modulus $E=(C^2_{11}-C^2_{12})/C_{11}$ . As shown in Table ~\ref{tab:ML_Young}, the Young moduli from different methods are close to each other. PBE and optB88-vdW are almost the same, while the results of LDA are larger than others. Although there are some numerical differences, the results of three methods present the same trends that Young modulus decrease with the atomic number of chalcogen and increase from Mo to W compounds. As shown above,  charge transfer  decreases and lattice constant increases in transition metal dichalcogenides MX$_2$   when X changes  from S to Te, which suggests that the reactivity and binding decrease from MS$_2$ to MTe$_2$.  The weakening binding between the metal and the chalcogen atom  thus leads to the smaller Young modulus. In addition, we can also find that the Young modulus for those monolayer TMDCs are comparable to graphene. For instance, young modulus for MoO$_2$ and WO$_2$ are almost the two-third of the graphene, which is relatively high. MoS$_2$, MoSe$_2$, WS$_2$ and WSe$_2$ have a value about one-third of the graphene showing excellent elastic property. The calculated Young modulus 141 Nm$^{-1}$ and 156 Nm$^{-1}$ using LDA for monolayer MoS$_2$ and WS$_2$ agree well with  theoretical studies $^{[13, 14]}$ and experiment measurement (180 $\pm$ 60 Nm$^{-1}$ $^{[9]}$ or around 170 Nm$^{-1}$ $^{[10]}$). We can also find that LDA calculations gave the closest results to experiment among three functionals, which suggest that LDA calculations are more accurate to describe the elastic property of monolayer TMDCs.
 \begin{table*}
\centering
\caption{\label{tab:ML_elastic}The calculated elastic constants of monolayer TMDCs in unit of  Nm$^{-1}$  calculated using PBE functional. The data of graphene is also listed for comparison.}
\begin{tabular}{ccccccccccccccccccccccccccccccccccccccccccccccc}
			\hline	
\hline																		
&	MoO$_2$	&	MoS$_2$	&	MoSe$_2$	&	MoTe$_2$	&	WO$_2$	&	WS$_2$	&	WSe$_2$	&	WTe$_2$	&	graphene	\\
\hline
	C$_{11}$	&	232.92	&	134.89	&	110.7	&	84.72	&	253.25	&	148.47	&	122.01	&	91.42	&	351.6	\\
	C$_{12}$	&	84.67	&	32.03	&	25.66	&	20.25	&	91.32	&	31.20	&	22.91	&	15.40&	64.10	\\
	C$_{44}$	&	74.12	&	51.43	&	42.52	&	32.23	&	80.96	&	58.63	&	43.03	&	38.01	&	143.75	\\
    $\nu$&	0.36	&	0.24	&	0.23	&	0.24	&	0.36	&	0.21	&	0.19	&	0.17	&	0.18	\\
\hline
\hline
\end{tabular}
\end{table*}

 \begin{table}
\centering
\caption{\label{tab:ML_Young}The Young modulus of TMDCs in unit of  Nm$^{-1}$  with PBE, LDA and vdW-DF(optB88-vdW) method.	 }
\begin{tabular}{ccccc}
\hline
\hline	
&	PBE	&	LDA	&	optB88	\\
\hline
MoO$_2$	&	202.1	&	221.8	&	199.8	\\
MoS$_2$	&	127.3	&	141.2	&	123.5	\\
MoSe$_2$	&	104.8	&	120.9	&	105.7	\\
MoTe$_2$	&	79.9	&	94.07	&	80.7	\\
WO$_2$	&	220.3	&	254.3	&	219.7	\\
WS$_2$	&	141.9	&	156.4	&	140.7	\\
WSe$_2$	&	114.4	&	132.8	&	118.2	\\
WTe$_2$	&	88.8	&	101.8	&	88.4	\\
\hline
\hline
\end{tabular}
\end{table}

\subsection{\label{sec:BL}  bilayer TMDCs}
 \subsubsection{\label{sec:bele}Structure properties}
 There are different configurations when monolayers are stacked into bilayer structure $^{[19]}$. As Fig.~\ref{fig:geom}-(c) shown, five stacking orders for the bilayer TMDCs are considered in this paper. Since the optB88-vdW can give a very similar in-plane lattice parameter with PBE and the  interlayer interaction is dominated by vdW forces, we only analysis the results of the bilayer TMDCs using optB88-vdW functional. And LDA calculation is also performed for comparison. The AA1 and AB with almost  degenerate energy are found to be the most stable configuration for all bilayer TMDCs, which also agrees with earlier calculations. $^{[19]}$ Firstly, we also give the structure parameters with five different stacking orders using LDA and optB88-vdW method in Table ~\ref{tab:BL_geom}. we can find that, the stack of monolayer does not change the monolayer structure much. The lattice constants of monolayer and bilayer TMDCs are almost the same for both LDA and optB88-vdW method, while the inter-layer distances represented as d$_{m-m}$ varies with the stacking order. Both  methods predict that the AA and AB1 stacking order have a larger value of d$_{m-m}$ than other stacking orders, while the distance for AA1 is the smallest, which agrees with the stability order of those configurations. The interlayer distances for bilayer TMDCs range from 5.2 to 7.8 \AA~ according to our calculation. It is clearly shown that the structural parameters including in-plane lattice parameter and inter-layer distance obtained from LDA are much smaller than that of optB88-vdW for the same bilayer structures.

 \begin{table*}  \scriptsize
\centering
\caption{\label{tab:BL_geom}The optimized structural parameter of  bilayer TMDCs using LDA and optB88-vdW method. a stand for lattice constant, d$_{m-m}$ is the distance of the metal atoms between different layers. }
\begin{tabular}{cccccccccccccccccccccccccc}
\hline
\hline
Stacking order	&		&	A1B	&		&	AA	&		&	AA1	&		&	AB	&		&	AB1	&		\\
	\hline
&		&	a(\AA)	&	d$_{m-m}$ (\AA)	&	a(\AA)	&	d$_{m-m}$ (\AA)	&	a(\AA)	&	d$_{m-m}$ (\AA)	&	a(\AA)	&	d$_{m-m}$ (\AA)	&	a(\AA)	&	d$_{m-m}$ (\AA)	\\
\hline
MoO$_2$	&	LDA	&	2.79	&	4.81	&	2.79	&	5.30	&	2.79	&	4.86	&	2.79	&	4.82	&	2.79	&	5.28	\\
	&	optB88	&	2.83	&	5.28	&	2.83	&	5.69	&	2.83	&	5.23	&	2.83	&	5.20	&	2.83	&	5.54	\\
MoS$_2$	&	LDA	&	3.12	&	6.09	&	3.12	&	6.80	&	3.12	&	6.01	&	3.12	&	5.98	&	3.12	&	6.68	\\
	&	optB88	&	3.18	&	6.36	&	3.19	&	6.75	&	3.19	&	6.28	&	3.19	&	6.24	&	3.19	&	6.71	\\
MoSe$_2$	&	LDA	&	3.25	&	6.52	&	3.25	&	7.05	&	3.25	&	6.33	&	3.25	&	6.35	&	3.25	&	7.05	\\
	&	optB88	&	3.33	&	6.64	&	3.32	&	7.14	&	3.32	&	6.57	&	3.33	&	6.57	&	3.32	&	7.09	\\
MoTe$_2$	&	LDA	&	3.47	&	7.08	&	3.46	&	7.73	&	3.47	&	6.85	&	3.47	&	6.89	&	3.47	&	7.71	\\
	&	optB88	&	3.56	&	7.25	&	3.56	&	7.78	&	3.56	&	7.10	&	3.56	&	7.11	&	3.56	&	7.75	\\
WO$_2$	&	LDA	&	2.79	&	4.88	&	2.79	&	5.34	&	2.79	&	4.90	&	2.79	&	4.87	&	2.79	&	5.31	\\
	&	optB88	&	2.83	&	5.27	&	2.83	&	5.68	&	2.83	&	5.22	&	2.83	&	5.19	&	2.83	&	5.58	\\
WS$_2$	&	LDA	&	3.13	&	6.18	&	3.13	&	6.81	&	3.13	&	6.06	&	3.13	&	6.04	&	3.12	&	6.69	\\
	&	optB88	&	3.18	&	6.37	&	3.19	&	6.82	&	3.19	&	6.29	&	3.19	&	6.26	&	3.19	&	6.73	\\
WSe$_2$	&	LDA	&	3.25	&	6.54	&	3.25	&	7.14	&	3.25	&	6.39	&	3.25	&	6.44	&	3.25	&	7.1	\\
	&	optB88	&	3.33	&	6.70	&	3.33	&	7.15	&	3.33	&	6.54	&	3.33	&	6.61	&	3.33	&	7.14	\\
WTe$_2$	&	LDA	&	3.47	&	7.13	&	3.47	&	7.76	&	3.47	&	6.87	&	3.47	&	6.95	&	3.47	&	7.72	\\
	&	optB88	&	3.57	&	7.27	&	3.56	&	7.79	&	3.56	&	7.12	&	3.57	&	7.12	&	3.56	&	7.76	\\
\hline
\hline
\end{tabular}
\end{table*}

  \subsubsection{\label{sec:ele}Electronic properties}

     \begin{figure}
\centering
\includegraphics[width=0.8\textwidth]{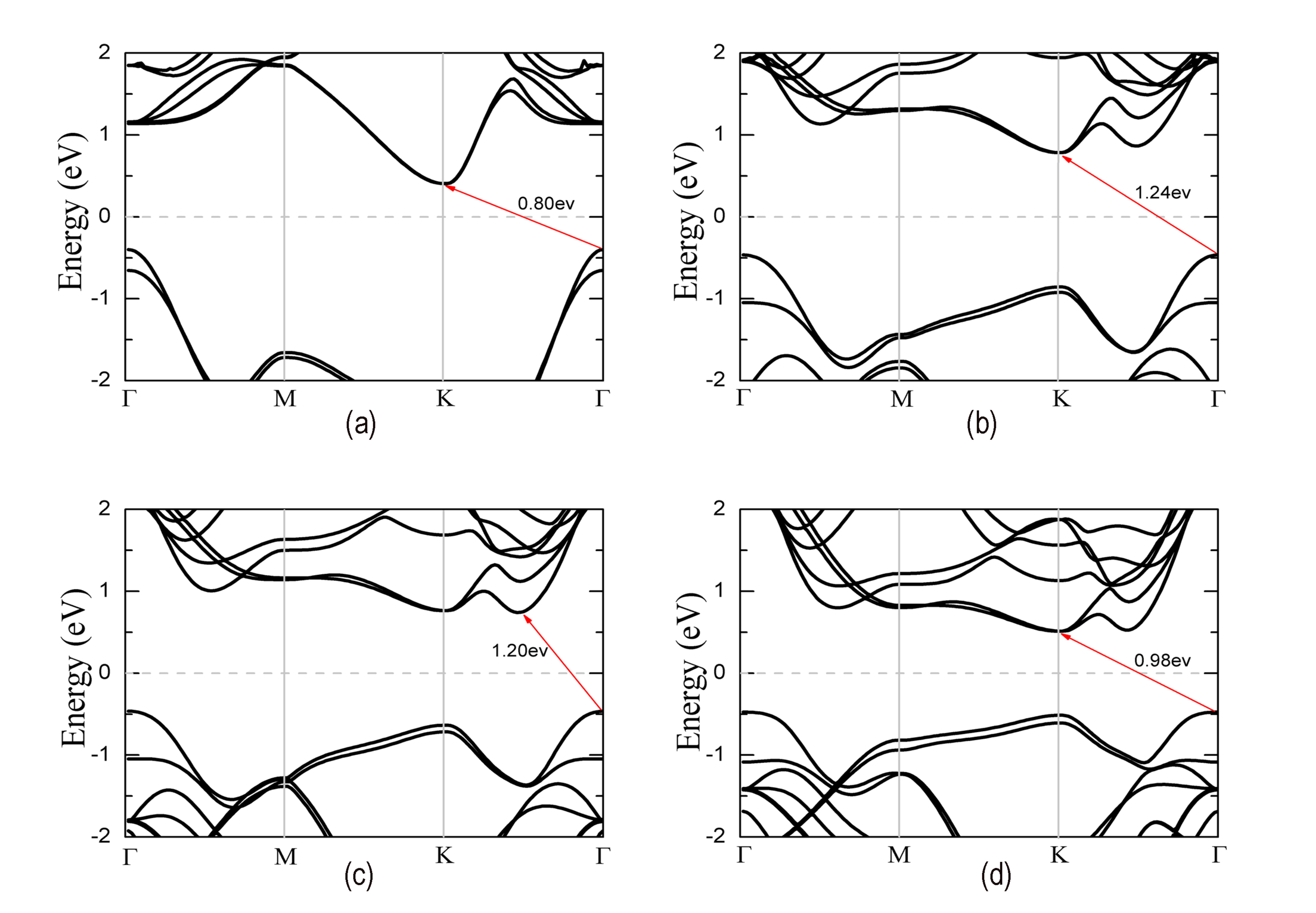}
\includegraphics[width=0.8\textwidth]{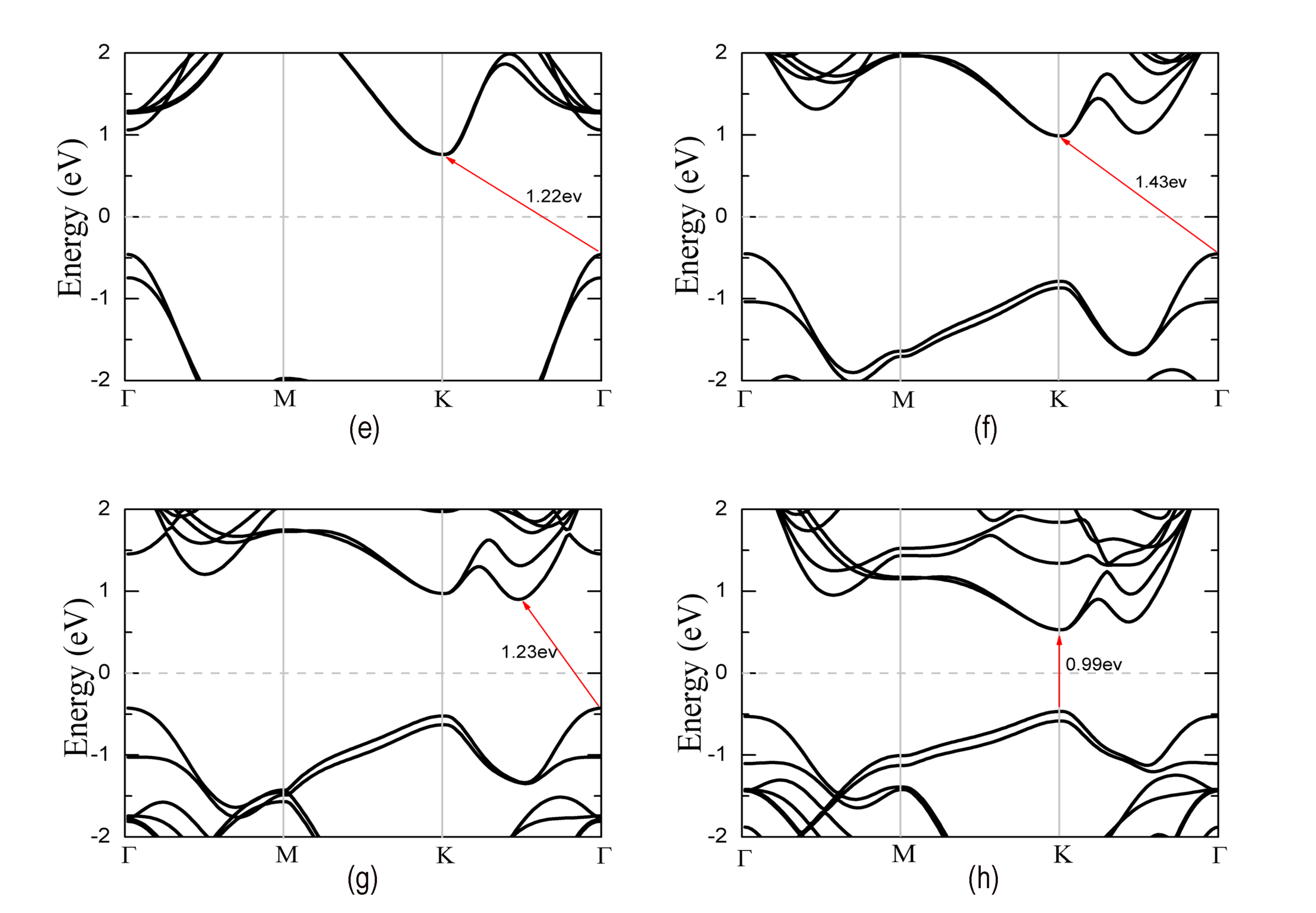}
\caption{\label{fig:BL_band}(Color online) The band structures for monolayer TMDCs calculated using PBE. (a)-(h) stand for MoO$_2$,MoS$_2$,MoSe$_2$,MoTe$_2$,WO$_2$,WS$_2$,WSe$_2$,WTe$_2$ respectively. The Fermi energy is marked with light gray dash lines. The band gaps are shown with red arrows. }
\end{figure}

The band gaps of TMDCs are known to be tunable with thickness $^{[5, 10, 29]}$. It is thus interest to investigate the electronic structure of bilayer TMDCs. Since the low-energy AA1 and AB are found to have similar  band structure, we only give the band structures of bilayer TMDCs in  AA1 stacking order in Fig.~\ref{fig:BL_band}. Unlike the direct band gap of monolayers, indirect band gap are found except for bilayer WTe$_2$ which still has a direct band gap at the K points as shown in Fig.~\ref{fig:BL_band}- (h). We can easily find that the band gaps have the same variation behaviour with the monolayers form MO$_2$ to MTe$_2$. It has to be mentioned that the valence-band maximum (VBM) of bilayer WTe$_2$ at K point is only 0.06 eV larger than the maximum energy for $\Gamma$ point, which is much smaller than energy difference in  other cases. When  monolayers  are stacked into bilayers , the band gaps were found  to decrease , which agree well with previous theoretical studies $^{[19, 30]}$.
  \subsubsection{\label{sec:ele}Elastic properties}

To reveal the evolution of mechanical properties of TMDCs with thickness, we also calculated the elastic constants, Young modulus and Poisson ratio of bilayer TMDCs in Table ~\ref{tab:BL_elastic}. The trend of bilayer Young moduli is very similar to monolayers. A decrease of Young modulus is found from MO$_2$ to MTe$_2$ and WX$_2$ are larger than MoX$_2$. The Young modulus of AA and AB1 stacking orders are found to be larger than others according to LDA. Meanwhile, we also find the differences of Young modulus for different stacking order are within 5\%, which indicates that the stacking order have little influence on the elastic properties of bilayer. Compared with optB88-vdW functional, LDA has a larger  Young modulus. Especially, LDA method predicts the Young moduli of around 280 Nm$^{-1}$ for bilayer MoS$_2$, which is consistent with recent experiment 300$\pm$13 Nm$^{-1}$$^{[10]}$. Table ~\ref{tab:BL_elastic} also shows the Poisson ratio of bilayers of TMDCs with different configurations. The calculated results show that these bilayers have close Poisson ratio to the corresponding monolayer. Among different TMDCs bilayer,  MoS$_2$, MoSe$_2$ and MoTe$_2$  have similar  Poisson ratio  with the value of  0.2, which is also independent on the stacking order. While  the Poisson ratios for bilayer WSe$_2$ and WTe$_2$, are found to be much smaller the others.  We also notice that  optB88-vdW predicts the  larger Poisson ratios  than LDA.
 \begin{table*}\footnotesize
\centering
\caption{\label{tab:BL_elastic} Young modulus and Poisson ratio for bilayer TMDCs with different stacking order. }
\begin{tabular}{cccccccccccccccccccccccccc}
\hline
\hline
 Young Modulus(Nm$^{-1}$)	&	A1B	&		&	AA	&		&	AA1	&		&	AB	&		&	AB1	&		\\
	&	LDA	&	optB88	&	LDA	&	optB88	&	LDA	&	optB88	&	LDA	&	optB88	&	LDA	&	optB88	\\
\hline
MoO$_2$	&	439.7	&	400.7	&	442.3	&	399.9	&	439.4	&	398.9	&	438.9	&	399.8	&	441.2	&	405.5	\\
MoS$_2$	&	281.9	&	255.7	&	285.9	&	249.9	&	282.0	&	252.2	&	276.4	&	254.0	&	283.2	&	248.2	\\
MoSe$_2$	&	236.9	&	210.4	&	241.1	&	213.4	&	238.2	&	212.8	&	236.1	&	212.2	&	240.7	&	213.7	\\
MoTe$_2$	&	182.9	&	163.0	&	189.0	&	167.7	&	186.6	&	162.0	&	182.1	&	158.6	&	186.8	&	161.7	\\
WO$_2$	&	503.4	&	461.9	&	509.1	&	470.8	&	504.2	&	471.1	&	503.9	&	468.9	&	509.6	&	461.9	\\
WS$_2$	&	314.1	&	291.7	&	313.2	&	279.6	&	309.7	&	282.2	&	309.9	&	280.7	&	313.7	&	282.3	\\
WSe$_2$	&	262.2	&	232.4	&	265.7	&	233.2	&	265.1	&	227.3	&	262.5	&	234.3	&	266.1	&	232.5	\\
WTe$_2$	&	198.8	&	172.5	&	204.1	&	177.9	&	203.2	&	175.9	&	200.1	&	172.3	&	202.9	&	176.9	\\
\hline
\\
\\
\hline
\hline
Poisson Ratio	&	A1B	&		&	AA	&		&	AA1	&		&	AB	&		&	AB1	&		\\
	&	LDA	&	optB88	&	LDA	&	optB88	&	LDA	&	optB88	&	LDA	&	optB88	&	LDA	&	optB88	\\
\hline
MoO$_2$	&	0.365	&	0.364	&	0.366	&	0.367	&	0.365	&	0.367	&	0.369	&	0.37	&	0.365	&	0.353	\\
MoS$_2$	&	0.212	&	0.239	&	0.199	&	0.253	&	0.227	&	0.247	&	0.218	&	0.234	&	0.222	&	0.256	\\
MoSe$_2$	&	0.201	&	0.224	&	0.205	&	0.232	&	0.214	&	0.230	&	0.211	&	0.223	&	0.209	&	0.229	\\
MoTe$_2$	&	0.202	&	0.221	&	0.210	&	0.218	&	0.213	&	0.242	&	0.217	&	0.248	&	0.212	&	0.245	\\
WO$_2$	&	0.342	&	0.348	&	0.337	&	0.331	&	0.336	&	0.329	&	0.341	&	0.332	&	0.340	&	0.348	\\
WS$_2$	&	0.180	&	0.194	&	0.195	&	0.220	&	0.201	&	0.219	&	0.198	&	0.213	&	0.197	&	0.216	\\
WSe$_2$	&	0.173	&	0.191	&	0.179	&	0.200	&	0.174	&	0.217	&	0.174	&	0.189	&	0.175	&	0.204	\\
WTe$_2$	&	0.154	&	0.193	&	0.155	&	0.188	&	0.159	&	0.199	&	0.160	&	0.201	&	0.158	&	0.189	\\
\hline
\hline
\end{tabular}
\end{table*}

It should be noticed that the inter-layer interaction is known to be predominated by vdW interaction, an accurate description for such a system may need a more advanced treatment such as RPA. However, on the basis of the reasonable description for elastic properties of TMDCs, LDA method may be seen as an alternative to describe the elastic properties of bilayer TMDCs. Recently, Liu \emph{et al.} $^{[10]}$have investigated the elastic modulus of CVD grown monolayer MoS$_2$ and WS$_2$ and their Bilayer Heterostructures and found that the 2D moduli of their bilayer heterostructures are lower than the sum of 2D modulus of each layer, which shows a strong mechanical inter-layer coupling between the layers. However, according to the present results, the Young modulus of bilayer structure calculated by both LDA and optB88-vdW is almost twice as that of the corresponding monolayer, which shows that the mechanical coupling between monolayers seems to be negligible in bilayers. Considered the experimental uncertainty,  further effort both in experimental and theory is thus suggested to investigate the mechanical coupling of monolayers in bilayer.
\section{\label{sec:conclusion} Conclusion}

We have investigated the band structure and elastic property for eight kinds of two dimensional TMDCs in monolayer and different bilayers using different functionals. Our results indicate that all those monolayers and bilayers belong to semiconductors and there is a transition from direct gaps for monolayers (except for MoO$_2$ and WO$_2$), into indirect gaps in bilayers (except WTe$_2$). The calculation also shows that the monolayer TMDCs  presents excellent elastic property, which are comparable to graphene. Meanwhile, we also find that the LDA method is an efficient alternative to more advanced method in describing the elastic property of both monolayer and bilayer TMDCs. Furthermore, our results also show that the elastic property is only influenced slightly by stacking order and the mechanical coupling between monolayers seems to be negligible.

\vspace*{2mm}



\begin{thebibliography}{99}
\itemsep=-4pt plus.2pt minus.2pt  
\small


\bibitem{1} Novoselov K S, Geim A K, Morozov S V, Jiang D, Zhang Y, Dubonos S V , Grigorieva I V and Firsov A A 2004 {\it Science} {\bf 306} 666
\bibitem{2}	Chen J H, Jang C, Xiao S, Ishigami M, and Fuhrer M S 2008 {\it Nature nanotechnology} {\bf 3} 206
\bibitem{3}	Lee C, Wei X, Kysar J W, and Hone J 2008 {\it Science} {\bf 321} 385
\bibitem{4}	Splendiani A , Sun L , Zhang Y , Li T , Kim J , Chim C Y and Galli G and Wang F 2010 {\it Nano letters} {\bf 10} 1271
\bibitem{5}	Mak K F, Lee C, Hone J, Shan J, and Heinz T F 2010 {\it Physical Review Letters} {\bf 105} 136805
\bibitem{6}	Wang Q H, Kalantar-Zadeh K, Kis A ,Coleman J N , and Strano M S 2012 {\it Nature nanotechnology} {\bf 7} 699
\bibitem{7}	Xu X, Yao W, Xiao D, and Heinz T F 2014 {\it Nature Physics} {\bf 10} 343
\bibitem{8}	Geim A K, and Grigorieva I V 2013 {\it Nature} {\bf 499} 419-425.
\bibitem{9}	Bertolazzi S, Brivio J and Kis A 2011 {\it ACS nano} {\bf 5} 9703
\bibitem{10} Liu K, Yan Q, Chen M, Fan W, Sun Y, Suh J, Fu D, Lee S, Zhou J, Tongay S, Ji J, Neaton J B and Wu J 2014 {\it Nano letters} {\bf 14} 5097
\bibitem{11} Ataca C, Topsakal M ,Akturk E And Ciraci S  2011 {\it The Journal of Physical Chemistry C}  {\bf 115} 16354
\bibitem{12} Cooper R C, Lee C, Marianetti C A, Wei X, Hone J, and Kysar J W, 2013  {\it Physical Review B} {\bf 87} 035423
\bibitem{13} Kang J, Tongay S, Zhou J, Li J and Wu J  2013 {\it Applied Physics Letters} {\bf 102} 012111
\bibitem{14} \c{C}akr D, Peeters F M, and Sevik C 2014 {\it Applied Physics Letters} {\bf 104} 203110
\bibitem{15} Wang Z-Y, Zhou Y-L, Wang X-Q, Wang F, Sun Q, Guo Z-X, Jia Y 2015 {\it Chinese Physics  B ) } {\bf 24} 026501
\bibitem{16} Molina-S\'anchez A, Sangalli D, Hummer K,Marini A and Wirtz L  2013 {\it Physical Review B} {\bf 88} 045412.
\bibitem{17} Yun W S, Han S W, Hong S C, Kim I G and Lee J D 2012 {\it Physical Review B} {\bf 85} 033305.
\bibitem{18} Ramasubramaniam A, Naveh D, and Towe E 2011 {\it Physical Review B} {\bf 84} 205325.
\bibitem{19} He J, Hummer K and Franchini C 2014 {\it Physical Review B} {\bf 89} 075409.
\bibitem{20} Kresse G and Hafner J 1993 {\it Physical Review B} {\bf 47} 558.
\bibitem{21} Kresse G and Furthm\"uller J 1996 {\it Physical Review B}  {\bf 54} 11169.
\bibitem{22} Klime\ifmmode \check{s}\else \v{s}\fi{} J, Bowler D R and Michaelides A 2010 {\it Journal of Physics: Condensed Matter} {\bf 22} 022201.
\bibitem{23} Klime\ifmmode \check{s}\else \v{s}\fi{} J, Bowler D R and Michaelides A 2011 {\it Physical Review B} {\bf 83} 195131.
\bibitem{24} Zhang W B ,Chen C and Tang P Y 2014 {\it The Journal of chemical physics} {\bf 141} 044708.
\bibitem{25} Cadelano E, Palla P L, Giordano S And Colombo L 2010 {\it Physical Review B} {\bf 82} 235414.
\bibitem{26} Ataca C and Ciraci S 2011 {\it The Journal of Physical Chemistry C} {\bf115} 13303
\bibitem{27} Shi H,Pan H,Zhang Y W and Yakobson B I 2013 {\it Physical Review B} {\bf 87} 155304
\bibitem{28} Castellanos-Gomez A, Poot M, Steele G A,van der Zant H S, Agra\"it N and Rubio-Bollinger G 2012 {\it Advanced Materials} {\bf 24} 772
\bibitem{29} Conley H J, Wang B, Ziegler J I, Haglund Jr R F, Pantelides S T and Bolotin K I 2013 {\it Nano letters} {\bf13} 3626
\bibitem{30} Ataca C, Sahin H, and Ciraci S 2012 {\it The Journal of Physical Chemistry C} {\bf116} 8983

\end{thebibliography}
\end{document}